\documentclass[11pt]{iopart}
\usepackage{graphicx,color}
\begin{document}

\title[Carbon
  nanotube sensor for vibrating molecules]{Carbon
  nanotube sensor for vibrating molecules}

\author{F Remaggi$^{1}$, N Traverso Ziani$^{1,2}$, G Dolcetto$^{1,2,3}$, F Cavaliere$^{1,2}$ and M Sassetti$^{1,2}$}
\address{$^{1}$ Dipartimento di Fisica, Universit\`a di Genova, Via
  Dodecaneso 33, 16146 Genova, Italy\\
\noindent $^{2}$ CNR-SPIN, Via Dodecaneso 33, 16146 Genova, Italy\\
\noindent $^{3}$ INFN, Via Dodecaneso 33, 16146 Genova, Italy}
\ead{fabio.cavaliere@gmail.com}
\begin{abstract}
  The transport properties of a CNT capacitively coupled to a molecule vibrating along one of its librational modes are studied and its transport properties analyzed in
  the presence of an STM tip. We evaluate the linear
  charge and thermal conductances of the system and its
  thermopower. They are dominated by {\em position-dependent} Franck-Condon factors, governed by a position-dependent effective coupling constant peaked at the molecule position. Both conductance and thermopower allow to extract some information on the position of the molecule along the CNT. Crucially,
  however, thermopower sheds also light on the vibrational
  levelspacing, allowing to obtain a more complete characterization of
  the molecule even in the linear regime.
\end{abstract}

\maketitle

\section{Introduction}
Carbon nanotubes (CNTs) are outstanding examples of one-dimensional
(1D) conductors~\cite{ijima,charlier}. When embedded into a circuit and
connected to external leads via tunneling barriers, CNTs behave as
quantum dots~\cite{kou,cobden,grint}. Owing to their one-dimensional nature, their
behaviour can be well described by means of a Luttinger liquid (LL)
model~\cite{voit,yoshi,grifoni,noi2,noi} which can also include
electron-electron interactions. The latter can lead to several
interesting effects, including the formation of Wigner
molecules~\cite{wm1,wm2,wm3}, peculiar correlated electron states
characterized by unusual transport properties~\cite{wm4} which in a
CNT can be either probed by means of
an {atomic force microscope (AFM)~\cite{wm5,wm6,wm7} or the tip of a scanning tunnel microscope (STM)}~\cite{wm8,wm9}. \\

\noindent Among the most prominent applications of CNTs is their use
as mass~\cite{mass1,mass2} or gas nanoscale
sensors~\cite{gas1,gas2}. Seminal work on gas sensing using CNT
detectors included the study of the change in resistance {of single wall nanotubes~\cite{kong} or mats} of
multiwall CNTs exposed to an environment containing H$_{2}$ or NH$_{3}
$ molecules~\cite{mat1,mat2}. The ultimate goal of CNT-based
nano-sensing is to achieve the resolution of few molecules and to be
able to identify the chemical species of the molecule itself. One
of the possible ``fingerprints'' of a molecule is its vibrational spectrum.\\

\noindent Molecules exhibit several different
vibrational modes, with a broad spectrum of vibrational
frequencies. The {\em intrinsic} vibrational modes strongly depend on
the molecule mass as well as its spatial orientation and on the nature
of the chemical bonds composing it. The
typical frequency of these modes can be as high as 100
THz (65 meV)~\cite{refvib}. Molecules coupled to a surface, such as the one of the CNT, show
additional {\em libration } vibrational modes involving the {\em
  center of mass} of the molecule itself which oscillates as a whole
with respect to the surface~\cite{witte}. Such modes are much softer,
with a vibrational frequency of the order of 1 THz and
excitation energies {\em smaller} than 1 meV.\\

\noindent Molecular vibrations are typically detected with optical
methods such as Raman scattering~\cite{refvib}. However, it is in
principle possible to employ the transport properties of a CNT also to
detect at least some of these vibrational modes. Low-energy modes have excitation energies smaller than the the typical
level spacing $\delta E$ of a CNT which, for a lentgh $L\approx 200\ \mathrm{nm}$ can be estimated as $\delta E=\hbar\pi v_{F}/L\approx
10$ meV ($v_{F}=8\cdot10^{5}\ \mathrm{m/s}$ the CNT Fermi velocity). This allows to single out their signatures in the transport properties, in contrast with the internal
vibration modes with excitation energies well exceeding $\delta E$, whose signatures are intertwined with those due to the electronic degrees of freedom of the CNT. In turns, this would lead to ``smart'' molecular sensors,
able to detect the presence and the type of
molecule attached to it.\\
\noindent This coupling of molecular vibrations and electronic states
in a CNT is strongly reminescent of the situation occurring in a
nano-electromechanical system {(NEMS)}~\cite{NEMS,flens,KochFC,KochFC2,Haupt,Merlo}, in which
mechanical and electronic degrees of freedom are strongly
coupled. Suspended carbon nanotubes themselves can behave as NEMS when
current flowing through them excites their vibrational
modes~\cite{noi,RapCom}.\\

\noindent Another powerful tool to detect the excitation spectrum of a
mesoscopic system is the thermopower~\cite{thermo}. In the context of
quantum dots, thermopower has been studied
theoretically~\cite{BeenakkerThermo,Turek,Kubala,Zhang,Costi,Liu,Milenow} as well as experimentally~\cite{Staring,Dzurak,Small,Scheibner}. It constitutes a powerful tool to investigate the spectrum of
excited states of quantum dots in the {\em linear regime}~\cite{BeenakkerThermo}, in contrast with the more conventional nonlinear transport spectroscopy. In
the context of NEMS, the thermopower has been shown experimentally to
provide a way to investigate mechanical vibrations in sequential as
well as in the cotunneling regime~\cite{Koch}. The information brought by thermopower has been employed in the study of molecular junctions~\cite{Segal,Zimbo}, in the presence of an STM tip~\cite{Krawiec}, and even in gas sensors based on CNT mats~\cite{eklund}.\\

\noindent In this paper we study the behaviour of a single-wall CNT capacitively coupled to a vibrating molecule. The CNT-molecule
system is coupled via tunnel barriers to a lateral contact and to a
STM tip free to scan the CNT length. Focusing on the linear regime, we evaluate the linear charge, the thermal conductance and the thermopower. Our task is to
investigate to which extent the linear transport properties bring
about information about the molecule attached to the CNT
surface. Our main reults are the following\\

\noindent 1. The capacitive coupling of the electrons and the molecule
leads to the appearance of {\em space-dependent}~\cite{noi,RapCom,PSCR} Franck-Condon (FC) factors~\cite{Franck,Condon}, whose behaviour is dominated by an effective, space-dependent CNT-molecule
coupling constant $\lambda(x)$ ($x$ the coordinate of the tip) which affects the tunneling rate through the tip and the lateral contact. Such
a coupling constant is peaked around the
position where the molecule sits;\\
\noindent 2. The linear conductance exhibits a maximum following a simple analytical function of $\lambda(x)$, allowing to identify where the molecule is by scanning the STM tip;\\
\noindent 3. The thermopower shows resonant features at energies
corresponding to the vibrational spectrum of the molecule, allowing a precise characterization of the latter.\\

\noindent Our findings show that thermopower measurements would be able to detect the
position of a molecule and to gain information about its vibrational
frequency.\\

\noindent The outline of the paper is as follows. In
Sec.~\ref{sec:sec1} the model for the CNT capacitively coupled to
the molecule is described and its spectrum diagonalized. In
Sec.~\ref{sec:sec2} a model
for the coupled system is introduced and its Hamiltonian is exactly
diagonalized by means of a canonical transformation. The setup for a
transport experiment is then presented, tunneling rates in the sequential regime are evaluated and
a master equation is
set-up. Finally, linear charge and thermal
conductances are obtained. Section~\ref{sec:sec3} contains the results
concerning the position-dependent coupling constant, linear
conductance and thermopower. Finally, in Sec.~\ref{sec:sec4} we
summarize our conclusions.
\section{Nanotube, molecule and coupling}
\label{sec:sec1}
\begin{figure}[htbp]
\begin{center}
\includegraphics[width=10cm,keepaspectratio]{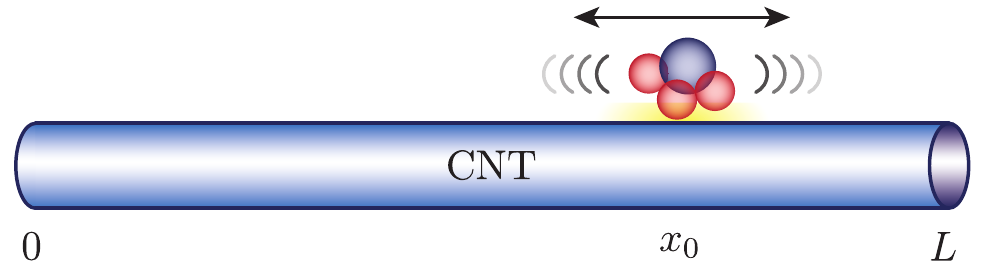}
\caption{Schematic rendering of the CNT coupled at $x_{0}$
  to a molecule vibrating along its libration mode.}
\label{fig:fig1}
\end{center}
\end{figure}
\noindent The device consists of a single-wall {metallic} CNT bounded in $x=0$ and $x=L$, capacitively coupled to a molecule
which vibrates around the position $x_{0}$. The system is schematically
depicted in Fig.~\ref{fig:fig1}. In the following we will concentrate
on the libration modes of the molecule, since
their excitation energy compares favorably with the energy scale at
which
transport experiments are performed.\\

\noindent The Hamiltonians of the decoupled CNT and vibrating molecule
are respectively ($\hbar=1$)
\begin{eqnarray}
H^{(0)}_{CNT}&=&\sum_{j}\left[\frac{E_{j}}{2}N_{j}^{2}+\sum_{q>0}\omega_{j}(q)a_{j,q}^{\dagger}a_{j,q}\right]\,
,\label{eq:H0cnt}\\
H^{(0)}_{mol}&=&{\frac{\mathbf{P}^{2}}{2M}+\mathcal{U}(\mathbf{R})}\, ,\label{eq:H0molpre}
\end{eqnarray}
{where $\mathbf{R}=(X,Z)$ and $\mathbf{P}=(P_{0},P_{z})$ and we neglect the motion of the molecule along the waist of the nanotube, i.e. we assume that the molecule fingerprint is larger than the CNT transverse dimension $\mathcal{D}\ll L$. Here, $Z$ denotes a radial coordinate perpendicular to the CNT axis.}\\

\noindent Equation~(\ref{eq:H0cnt}) represents the bosonized Hamiltonian of the
CNT within the LL language~\cite{yoshi,grifoni,noi}. It is composed of
four sectors $j\in\{c+,c-,s+,s-\}$ where $c+$ ($s+$) represent the
total charge (spin) modes and $c-$ ($s-$) represent the relative
charge (spin) ones. The first term represents the
contribution of the CNT zero modes, with $N_{j}$ the number of extra
electrons per sector with respect to a neutral reference situation
with $N_{0}$ electrons. In the following, for
definiteness, we assume $N_{0}$ to be a multiple of four and that each
sector $j$ thus contains $N_{0}/4$ electrons. We have introduced
{$E_{j}=\pi v_{j}/4Lg_{j}$} with $v_{j}=v_{F}/g_{j}$ the propagation
velocity of the mode $j$ and $g_{j}$ the LL interaction parameters,
which capture the effects of short-range forward scattering among the
electrons. One has $g_{c+}=g\leq 1$ with $g=1$ for noninteracting
electrons. All other parameters $g_{j\neq c_{+}}=1$ due to symmetries
of the model~\cite{yoshi}. The second term represents {\em collective}
excitations of bosonic character, with spectrum $\omega_{j}(q)=v_{j}q$ with $q=\pi n_{q}/L$ the quantized momentum ($n_{q}>0$ an integer) and canonical bosonic operators $a_{j,q}$.\\

\begin{figure}[htbp]
\begin{center}
\includegraphics[width=10cm,keepaspectratio]{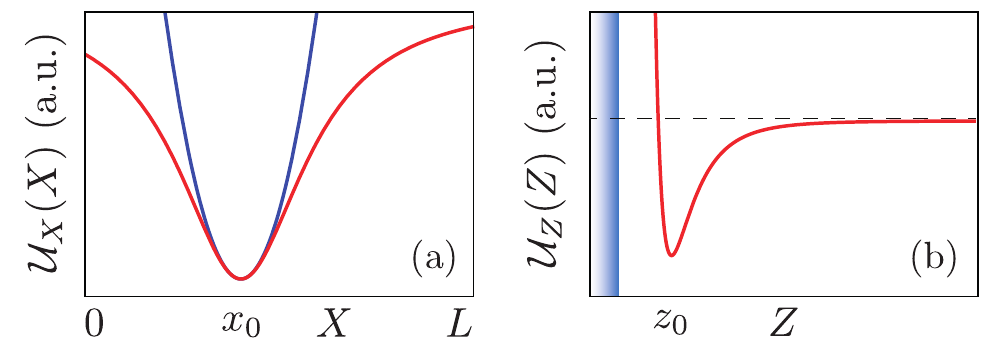}
{\caption{Schematic depiction of the potential wells where the molecule sits (a) along the nanotube and (b) in the radial direction. The blue curve in panel (a) represents the harmonic approximation assumed in this work. The shaded portion in panel (b) suggests the location of the CNT surface.}}
\label{fig:fig2}
\end{center}
\end{figure}

\noindent {The molecule sits in a potential well which, we assume separable $\mathcal{U}(\mathbf{R})=\mathcal{U}_{X}(X)+\mathcal{U}_{Z}(Z)$. The shape of $\mathcal{U}_{X}(X)$ and $\mathcal{U}_{Z}(Z)$ is schematically depicted in Fig.~\ref{fig:fig2}. The molecule is located around the position $x_{0}$, corresponding to the minimum of $\mathcal{U}_{X}(X)$. A rather shallow well is expected, which can be very well approximated by a harmonic potential $\mathcal{U}_{X}(X)\sim\frac{M\omega_{0}^{2}}{2}(X-x_0)^2$. Concerning the $Z$ direction, the molecule occupies the minimum of a potential well constituted by a repulsive short-range part and an attractive long range tail. Such well, located at a position $z_{0}$, is in general much narrower than the one which develops along the $X$ direction~\cite{hofmann}. As  a consequence, vibrations in the transverse, $Z$ direction have typical frequencies $\omega_{Z}\gg\omega_{0}$. In this work, we concentrate on linear transport properties and low temperatures $k_{B}T\ll\omega_{Z}$, so that the motion in the $Z$ direction is effectively "frozen" in the ground state of $\mathcal{U}_{Z}(Z)$. Therefore, upon a redefinition of the zero of energies, the Hamiltonian of the molecule simplifies as
\begin{equation}
H^{(0)}_{mol}=\frac{P_{0}^{2}}{2M}+\frac{M\omega_{0}^{2}}{2}(X-x_{0})^2\, ,\label{eq:H0mol}
\end{equation}
}

\noindent The molecular vibrations around $x_{0}$ are thus described by a simple harmonic
oscillator in Eq.~(\ref{eq:H0mol}), with $M$ the molecule mass, $X=x_{0}+X_{0}$ the fluctuating position of the molecule and $X_{0}$, $P_{0}$ the molecule displacement and momentum operators
respectively. Here and in the following, we will consider the case of vibration frequencies $\omega_{0}<\omega_{1}$ with 
\begin{equation}
\omega_{1}=\frac{\pi v_{F}}{gL}\, .
\end{equation}

\noindent The molecule and the CNT are assumed to be electrostatically
coupled via the term
\begin{equation}
H_{mix}=\int_{0}^{L}\ \mathrm{d}x\ V(x-X)\rho(x)\,,\label{eq:coupl1}
\end{equation}
where $V(x)$ represents the coupling potential and $\rho(x)$ the total CNT
electron density. {In this work, we neglect
charge tunneling between the CNT and the molecule, which may be the subject of future investigations. We assume that the molecule is physisorbed to the CNT, neglecting chemisorption effects which may induce structural modifications of the local nature of chemical bonds~\cite{chang,goldoni}.\\
\noindent Here, $V(x)$ represents the electrostatic potential induced on the CNT conduction electrons by the molecule charge cloud.}  As a model, we choose a {finite-range} screened potential
\begin{equation}
V(x)=\frac{V_{0}}{2w}e^{-\frac{|x|}{w}}\,,\label{eq:yukawa}
\end{equation}
with $V_{0}>0$ describing the amplitude of the potential and $w$ the interaction range.  {It must be stressed however that all the results presented in this paper do not depend qualitatively on the particular choice of $V(x)$ as long as it is peaked with a width of the order $w$.} Assuming vibrations of {\em
  small} amplitude, we expand Eq.~(\ref{eq:coupl1}) to lowest order
and obtain
\begin{equation}
H_{mix}\approx\int_{0}^{L}\ \mathrm{d}x\
V(x-x_{0})\rho(x)-X_{0}\int_{0}^{L}\
\mathrm{d}x\ \rho(x)\partial_{x}V(x-x_{0})\label{eq:coupl2}\, .
\end{equation}
\noindent The electron density can be decomposed as the sum
of terms $\rho(x)=\sum_{l\geq 0}\rho^{(l)}(x)$, representing the
long-wave term ($l=0$) and spatially oscillating contributions with
wavelengths $\ell_{l}\approx(2lk_{F})^{-1}$ with $k_{F}=\pi N_{0}/4L$
the CNT Fermi momentum. For $l=1$ one has Friedel oscillations, due to
finite-size effects~\cite{fabrizio}. Terms with $l>1$ are induced for
instance by electron-electron correlations, band curvature effects or
scattering with impurities which may lead to the formation of a Wigner
molecule~\cite{wm6,wm7,wm9,Wiglut}. In
typical CNT sensors, one has $N_{0}\gg 1$, with correspondingly short
wavelengths: in particular one can expect that already the Friedel wavelength $\ell_{1}\ll w$ for a
typical value of $N_{0}\approx 100$. In this case, the rapid
oscillations of terms with $l\geq 1$ give a vanishing contribution in
Eq.~(\ref{eq:coupl2}). {We are assuming here that the molecule does not induce appreciable back-scattering on the CNT conduction electrons~\cite{adessi,choi}. These effects may be the subject of future investigations.} Only the {\em long-wave} term of the electron density survives, which reads
\begin{equation}
\rho(x)=\rho^{(0)}(x)=e\frac{N_{c+}}{L}+\frac{e}{2\pi}\left[\partial_{x}\phi_{c+}(x)+x\to-x\right]\label{eq:lw1}
\end{equation}
with $e$ the electron charge and
\begin{equation}
\phi_{j}(x)=\sum_{q>0}\sqrt{\frac{\pi}{qL}}\left\{\frac{1}{\sqrt{g_{j}}}\cos{(qx)}\mathcal{A}_{j,q}^{(+)}+i\sqrt{g}\sin{(qx)}\mathcal{A}_{j,q}^{(-)}\right\}e^{-\alpha
      q/2}\, ,\label{eq:fop}
\end{equation}
LL field operators with $\mathcal{A}_{j,q}^{(\pm)}=a_{j,q}\pm
a_{j,q}^{\dagger}$ and $\alpha=k_{F}^{-1}$ a short-length cutoff. As it is clear,
the coupling
involves the total charge sector $j=c+$ only.\\
\noindent To proceed, we introduce $\tilde{X}_{\mu}=i\sqrt{Lg/2\pi
  v_{F}\mu}\left(a_{c+,\pi\mu/L}-a_{c+,\pi\mu/L}^{\dagger}\right)$
with $\mu\geq 1$, and the canonically conjugated momenta
$\tilde{P}_{\mu}=\sqrt{\pi
  v_{F}\mu/2gL}\left(a_{c+,\pi\mu/L}+a_{c+,\pi\mu/L}^{\dagger}\right)$. The charge sector $j=c+$ of Eq.~(\ref{eq:H0cnt}) becomes
of $\tilde{X}_{\mu}$ and $\tilde{P}_{\mu}$ 
\begin{equation}
H^{(0)}_{c+}=\frac{E_{c+}}{2}N_{c+}^{2}+\sum_{\mu\geq
  1}\left(\frac{\tilde{P}_{\mu}^{2}}{2}+\omega_{\mu}^{2}\frac{\tilde{X}_{\mu}^{2}}{2}\right)\,.\label{eq:H0cnt2}
\end{equation}
where $\omega_{\mu}=\pi\mu v_{F}/Lg$. Introducing
$\tilde{X}_{0}=\sqrt{M}X_{0}$ and $\tilde{P}_{0}=P_{0}/\sqrt{M}$ the
Hamiltonian of molecular vibrations becomes
\begin{equation}
H^{(0)}_{mol}=\frac{\tilde{P}_{0}^{2}}{2}+\omega_{0}^{2}\frac{\tilde{X}_{0}^{2}}{2}\, .\label{eq:H0mol2}
\end{equation}
The coupling in Eq.~(\ref{eq:coupl2}) can be decomposed into three terms
$H_{mix}=H_{mix}^{(1)}+H_{mix}^{(2)}+H_{mix}^{(3)}$ as
\begin{eqnarray}
H_{mix}^{(1)}&=&\gamma_{1}N_{c+}\, ,\label{eq:coup1}\\
H_{mix}^{(2)}&=&\sum_{\mu\geq
  1}\gamma_{2}(\mu)\tilde{X}_{\mu}\, ,\label{eq:coup2}\\
H_{mix}^{(3)}&=&\sum_{\mu\geq 1} \gamma_{3}(\mu)\tilde{X}_{0}\tilde{X}_{\mu}\, ,\label{eq:coup3}\\
\end{eqnarray}
where
\begin{eqnarray}
\gamma_{1}&=&\xi\omega_{0}\frac{L}{\ell_{0}}\mathcal{I}_{1}\, ,\\
\gamma_{2}(\mu)&=&\xi\omega_{1}^{2}\sqrt{2\eta^{3}}\frac{L}{\ell_{0}}\mathcal{I}_{2}(\mu)\, ,\\
\gamma_{3}(\mu)&=&-\pi\xi\omega_{1}^{2}\sqrt{2\eta^{3}}\mathcal{I}_{3}(\mu)\, ,\label{eq:gamma3}
\end{eqnarray}
and
\begin{eqnarray}
\mathcal{I}_{1}&=&\frac{1}{2w}\int_{0}^{L}{\mathrm{d}}x\ e^{-\frac{|x-x_{0}|}{w}}\, ,\\
\mathcal{I}_{2}(\mu)&=&\frac{\mu
  e^{-\frac{\alpha\mu}{2L}}}{2w}\int_{0}^{L}\mathrm{d}x\ e^{-\frac{|x-x_{0}|}{w}}\sin{\left(\frac{\pi\mu x}{L}\right)}\, ,\\
\mathcal{I}_{3}(\mu)&=&\frac{\mu^{2} e^{-\frac{\alpha\mu}{2L}}}{w}\int_{0}^{L}\mathrm{d}x\ e^{-\frac{|x-x_{0}|}{2w}}\cos{\left(\frac{\pi\mu x}{L}\right)}\, .
\end{eqnarray}
Here, we have introduced the dimensionless parameters
\begin{eqnarray}
\xi&=&\frac{eV_{0}}{L\omega_{0}}\frac{\ell_{0}}{L}\label{eq:xidef}\, ,\\
\eta&=&\frac{\omega_{0}}{\omega_{1}}\label{eq:etadef}\, ,
\end{eqnarray}
with {$\ell_{0}=\left(M\omega_{0}\right)^{-1/2}$} the characteristic oscillator length.\\
\noindent The terms in Eq.~(\ref{eq:coup1}) and Eq.~(\ref{eq:coup2}) can be
promptly eliminated by means of a linear shift of the operators
$N_{c+}$ and $X_{\mu}$. The shift of $N_{c+}$ in Eq.~(\ref{eq:coup1}) leads to an offset of the gate potential considered in transport. The linear shift of $X_{\mu}$ produces a
constant energy shift and hence its effects will be irrelevant on
transport properties where only energy differences come into play. We
can therefore discard terms in Eq.~(\ref{eq:coup1}) and Eq.~(\ref{eq:coup2})
and focus on Eq.~(\ref{eq:coup3}), which couples the vibrational mode
$\tilde{X}_{0}$ only to the plasmonic excitations
$\tilde{X}_{\mu}$. The bosonic part of Eq.~(\ref{eq:H0cnt2}), Eq.~(\ref{eq:H0mol2}) and Eq.~(\ref{eq:coup3}) constitute a quadratic form which can be
diagonalized by means of a canonical transformation~\cite{Ullersma,noi} 
\begin{equation}
\tilde{P}_{\mu}=\sum_{\nu\geq 0}k_{\mu\nu}\tilde{p}_{\nu}\ ;\
\tilde{X}_{\mu}=\sum_{\nu\geq 0}k_{\mu\nu}\tilde{x}_{\nu}\, .\label{eq:canonical}
\end{equation}
The diagonalization procedure relies on the condition $\omega_{0}<\omega_{1}$~\cite{Ullersma}, i.e. $\eta<1$. This is satisfied if $L<L^{*}$ with $L^{*}=\pi v_{F}/g\omega_{0}$. For a typical value $\omega_{0}\approx 1\ {\mathrm{THz}}$ one obtains $L^{*}\approx 2.5\ \mu\mathrm{m}$.\\
\noindent The diagonalized Hamiltonian is
\begin{equation}
H_{mol/CNT}=\sum_{\mu\geq 0}\left(\frac{\tilde{p}_{\mu}^{2}}{2}+\Omega_{\mu}^{2}\frac{\tilde{x}_{\mu}^{2}}{2}\right)\, ,
\end{equation}
written in terms of the new collective normal modes
$\{\tilde{x}_{\mu},\tilde{p}_{\mu}\}$ with energy $\Omega_{\mu}$, which exhibit a
hybrid vibrational/plasmonic character. The coefficients $k_{\mu\nu}$
are given by~\cite{Ullersma}
\begin{equation}
k_{\mu\nu}=\frac{\gamma_{3}(\mu)}{\Omega_{\nu}^{2}-\omega_{\mu}^{2}}\left[1+\sum_{\mu\geq
  1}\frac{\gamma_{3}^{2}(\mu)}{\left(\Omega_{\nu}^{2}-\omega_{\mu}^{2}\right)^{2}}\right]^{-1/2}\, ,
\end{equation}
while the eigenenergies $\Omega_{\mu}$ are obtained solving the self-consistent equation
\begin{equation}
\Omega_{\mu}^{2}=\omega_{0}^{2}+\sum_{\nu\geq
  1}\frac{\gamma_{3}^{2}(\nu)}{\Omega_{\mu}^{2}-\nu^{2}\omega_{1}^{2}}\, .\label{eq:sceq}
\end{equation}
The condition
\begin{equation}
	\frac{1}{2\pi^{2}\xi^{2}\eta}\leq\sum_{\nu\geq 1}\frac{\mathcal{I}_{3}^{2}(\nu)}{\nu^{2}}\label{eq:nowb}
\end{equation}
ensures the positivity of the solutions of Eq.~(\ref{eq:sceq}) preventing the Wentzel-Bardeen instability~\cite{Ullersma}.

\noindent As a general feature, the mode with $\mu=0$ is reminescent
of a vibrational mode dressed by a polaron cloud with energy
$\Omega_{0}<\omega_{0}$. As we will see, this polaron cloud has profound consequences on the transport properties of the system. All other modes with $\mu\geq 1$ are
plasmon-like with a slightly increased energy
$\mu\omega_{1}<\Omega_{\mu}<(\mu+1)\omega_{1}$ for $\mu\geq1$.\\
For reasonable values of the system parameters (see Sec.~\ref{sec:sec3} for further details) $\xi=0.1$, $\eta=0.6$, $w=L/10$ and $\alpha=L/100$, one finds $\Omega_{0}\approx0.6\omega_{0}$ and $\Omega_{\mu\geq 1}\approx \mu\omega_{1}$.\\

\noindent Due to the diagonalization, also the operator $\phi_{c+}(x)$ is modified. In the low energy sector, we are only interested into the vibration-like excitations, with an energy of the order of $\Omega_{0}$. We can thus disregard the electronic collective excitations and approximate 
{
\begin{equation}
\phi_{c+}(x)\approx\sum_{\nu\geq
  1}k_{\nu0}\left[\sqrt{2g\omega_{1}}\sin\left(\frac{\pi\nu
      x}{L}\right)X_{0}+\sqrt{\frac{2}{\nu^2 g\omega_{1}}}\cos\left(\frac{\pi\nu
      x}{L}\right)P_{0}\right]\, .\label{eq:phicplus}
\end{equation}
}
\section{STM tip coupling and transport properties}
\label{sec:sec2}
\begin{figure}[htbp]
\begin{center}
\includegraphics[width=10cm,keepaspectratio]{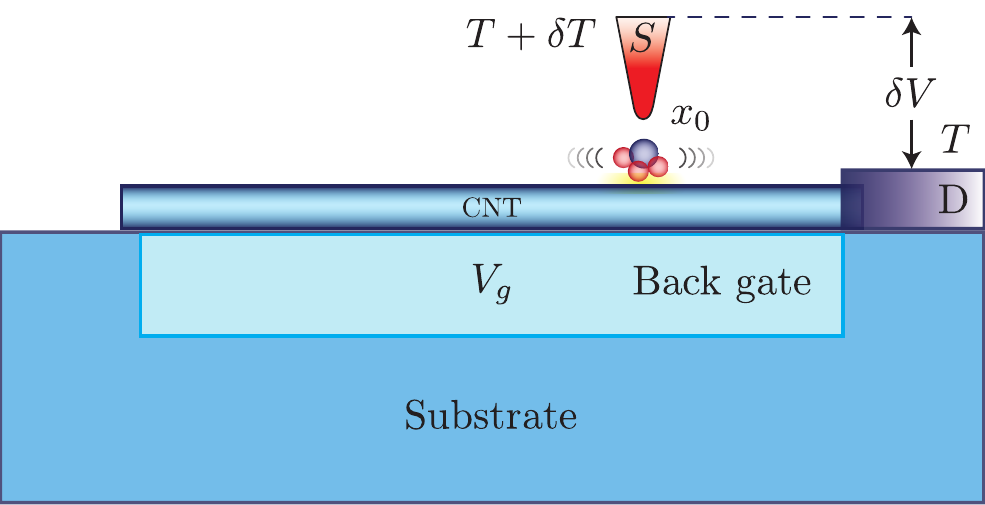}
\caption{Setup of the proposed STM-CNT molecular sensor. The CNT is
  tunnel-coupled to a STM tip which is able to scan the whole tube
  length, and to a lateral metallic contact. A back-gate, at potential $V_{g}$, is capacitively coupled to the
  CNT. The tip is kept at a temperature $T+\delta T$ while contacts
  are at a temperature $T$. See text for further details.}
\label{fig:fig3}
\end{center}
\end{figure}
Let us now turn to a setup which allows to investigate the transport
properties. It is schematically depicted in
Fig.~\ref{fig:fig3}. The CNT-molecule system is tunnel-coupled to a
lateral metallic contact and a STM tip, labeled respectively by the
index $\eta=D,S$, with tunneling amplitudes $\tau_{\eta}$. The  tip
scans the entire CNT length. An additional gate,
capacitively coupled to the CNT, is kept at a potential $V_{g}$ and
allows to tune the CNT chemical potential.\\

\noindent A promising tool to extract information about the spectrum of
molecular excitations even in the linear regime is the {\em
  thermopower}. Suppose that the lateral contacts are kept at a temperature
$T$ while the tip is placed at $T+\delta T$, with $\delta T\ll T$. As
a consequence of the temperature gradient, electrons flow through the
CNT-molecule system. The tip and the lateral leads are kept in an
open-circuit configuration, which implies a steady current $I=0$. As a result of the redistribution of
eletrons, a voltage drop $\delta V$
between the tip and the contacts develops. Eventually, an {\em
  equilibrium} situation is reached. {The {\em thermopower} $S$ is defined as
\begin{equation}
S=-\frac{\delta V}{\delta T}\,  .
\end{equation}
In the linear regime, the equilibrium condition
\begin{equation}
I\approx G_{V}\delta V+G_{T}\delta T=0\, ,
\end{equation}
with
\begin{equation}
G_{V}=\frac{\partial I}{\partial V}\,;\,G_{T}=\frac{\partial
  I}{\partial T}\, ,
\end{equation}
the charge and thermal linear conductances of the system, allows to obtain
\begin{equation}
S=\frac{G_{T}}{G_{V}}\, .
\end{equation}}
\noindent We will consider the sequential tunneling regime, assuming that the
typical tunneling rate throught the STM vacuum barrier or the leads
barrier is large with respect to the average rate of electrons
flow. This is valid when the temperature $T>\Gamma$, the
average level broadening due to tunneling, and the system is not in the deep Coulomb blockade regime. However, it has been shown that for the low-temperature regime $\Gamma<T<k_{B}^{-1}\omega_{0}$, both elastic and inelastic cotunneling contributions to the termopower have negligible effects even in the linear regime away from Coulomb oscillation resonances~\cite{Koch,Turek}.\\

\noindent A convenient tool to tackle the transport properties in
the sequential regime is the master equation for the
occupation probabilities of the CNT-molecule system states
$\mathcal{S}$. As already stated, we are interested in energies {\em smaller} than $\omega_{1}$, thus retaining
only the vibration-like excitations. The CNT-molecule
Hamiltonian can thus be written as
\begin{equation}
  H_{CNT-mol}=\frac{E_{c+}}{2}\left(N_{c+}-N_{g}\right)^2+\sum_{j\neq
    c_{+}}\frac{E_{j}}{2}N_{j}^{2}+\frac{\tilde{p}_{0}^{2}}{2}+\Omega_{0}^{2}\frac{\tilde{x}_{0}^{2}}{2}\, .\label{eq:hamham}
\end{equation}
{Here we have included into the CNT-molecule Hamiltonian the effects of the gate kept at a potential $V_{g}$, where $N_{g}=C_{g}V_{g}/e$ and $C_{g}$ the gate capacitance}.\\
\noindent The contacts and STM tip are treated as noninteracting Fermi
gases. The tunneling Hamiltonian which describes the connection
between the CNT and the contacts is $H_{t}=\sum_{r}H_{t}^{(r)}$ ($r=S,D$)
with~\cite{noi,ber1,ber2}
\begin{eqnarray}
H_{t}^{(S)}&=&\tau_{S}\sum_{\alpha,s}\Psi_{\alpha,s}^{\dagger}(x)\psi_{s,F}(0^{+})+\mathrm{h.c.}\,
,\\
H_{t}^{(D)}&=&\tau_{D}\sum_{\alpha,s,q}\Psi_{\alpha,s}^{\dagger}(x_{D})c_{s,q}+\mathrm{h.c.}\,
.
\end{eqnarray}
where $x_{D}=L$ is the position of the lateral contact. Here,
$\Psi_{\alpha,s}$ is the field operator for {\em right-movers} in the
CNT, with Dirac valley index $\alpha$ and spin $s$. It is given
by~\cite{noi}
\begin{equation}
\psi_{\alpha,s}(x)=\frac{J_{\alpha,s}}{\sqrt{2\pi\alpha}}e^{-i\Theta_{\alpha,s}}e^{\frac{i\pi\mathcal{N}_{\alpha,s}x}{4L}}e^{\frac{i\Phi_{\alpha,s}(x)}{2}}\, ,
\end{equation}
where $J_{\alpha,s}$ is a Klein factor and
\begin{eqnarray}
\Theta_{\alpha,s}&=&\theta_{c+}+\alpha\theta_{c-}+s\theta_{s+}+s\alpha\theta_{s-}\\
\mathcal{N}_{\alpha,s}&=&N_{c+}+\alpha N_{c-}+sN_{s+}+s\alpha N_{s-}\\
\Phi_{\alpha,s}(x)&=&\phi_{c+}(x)+\alpha \phi_{c-}(x)+s\phi_{s+}(x)+s\alpha \phi_{s-}(x)
\end{eqnarray}
with $[\theta_{j},N_{j'}]=i\delta_{j,j'}$. The operator
$\psi_{s,F}(z)$ describes the forward propagating modes in the STM
tip, with $z$ the vertical coordinate within the tip and $z=0^{+}$ the
tip vertex. Finally, $c_{s,q}$ are standard Fermi operators describing
electrons with spin $s$ and momentum $q$ in the lateral contact
$\eta=D$.\\

\noindent Due to the spin-valley simmetry of a CNT, degeneracies occur in the spectrum
of the Hamiltonian Eq.~(\ref{eq:hamham}). In particular, we will
consider in the following the resonance between states with
$N_{c+}=4\kappa$ (filled shell, non degenerate) and $N_{c+}=4\kappa+1$
electrons, which is fourfold degenerate~\cite{noi}. We will
label the CNT-molecule states as $\mathcal{S}=|N_{c+},l\rangle$ where
$l$ is the vibrational quantum number, and keep track of the
degeneracy via the factors $D_{\mathcal{S}}$. {We note here that even though our calculations have been performed for a CNT with a full fourfold degeneracy, the model can promptly take in account chirality effects leading to the emergence of a lower, twofold degeneracy due to the generality of the factors $D_{S}$. No qualitative modification of the results presented in the rest of the paper are expected in this case.} The occupation
probabilities $P_{\mathcal{S}}$ of the system states are thus found,
in the steady state, solving the master equation~\cite{master}
\begin{equation}
-P_{\mathcal{S}}\sum_{\mathcal{S}'\neq\mathcal{S}}D_{\mathcal{S}'}\Gamma_{\mathcal{S}\to\mathcal{S}'}(x)+\sum_{\mathcal{S}'\neq\mathcal{S}}P_{\mathcal{S}'}D_{\mathcal{S}}\Gamma_{\mathcal{S}'\to\mathcal{S}}(x)=0\label{eq:master}
\end{equation}
with the normalization condition
$\sum_{\mathcal{S}}P_{\mathcal{S}}=1$. Here, the tunneling rates are
\begin{equation}
\Gamma_{\mathcal{S}\to\mathcal{S}'}(x)=\Gamma_{\mathcal{S}\to\mathcal{S}'}^{(D)}+\Gamma_{\mathcal{S}\to\mathcal{S}'}^{(S)}(x)
\end{equation}
where $x$ is the STM tip position along the CNT. The steady current can be conveniently evaluated on either the tip or the lateral contact. One finds
\begin{equation}
I=e\sum_{l,l'}\left(P_{|4\kappa,l\rangle}\Gamma_{|4\kappa,l\rangle\to|4\kappa+1,l'\rangle}^{(S)}-P_{|4\kappa+1,l\rangle}\Gamma_{|4\kappa+1,l\rangle\to|4\kappa,l'\rangle}^{(S)}\right)\, .\label{eq:current}
\end{equation}

\noindent The tunneling rate for a transition between the initial state $|I\rangle$ and the final state $|F\rangle$ of the whole system - including the CNT, the STM tip and the lateral contact - is given by the Fermi golden rule
\begin{equation}
	\Gamma_{I\to F}=2\pi\left|\langle F|H_{t}|I\rangle\right|^2\delta(E_{F}-E_{I})\, ,\label{eq:grule}
\end{equation}
where $E_{I}$, $E_{F}$ are the energies of the whole system in the initial and final state respectively. The calculation proceeds performing a trace of Eq.~(\ref{eq:grule}) over the degrees of freedom of the tip and of the lateral contact, assumed in thermal equilibrium, in contrast with the charge degree of freedom and the occupation number of the vibrational state which retain their full dynamics. We will omit here the details of this standard procedure and directly quote the final results~\cite{RapCom}
considering for simplicity the the case of tunneling into the CNT only
\begin{eqnarray}
\Gamma_{|4\kappa,l\rangle\to|4k+1,l'\rangle}^{(S)}(x)&=&\Gamma_{0}^{(S)}F_{ll'}(x)f_{S}\left(\Delta
\bar{\epsilon}+\Omega_{ll'}+e\frac{\delta V}{2}\right)\label{eq:R1}\\
\Gamma_{|4\kappa,l\rangle\to|4k+1,l'\rangle}^{(D)}&=&\Gamma_{0}^{(D)}F_{ll'}(x_{D})f_{D}\left(\bar{\epsilon}+\Omega_{ll'}-e\frac{\delta V}{2}\right)\label{eq:R2}
\end{eqnarray}
where $\Gamma_{0}^{(\eta)}=2\pi\nu_{\eta}|\tau_{\eta}|^{2}$ with
$\nu_{\eta}$ the density of states of lead $r$. Here
\begin{equation}
	F_{ll'}(x)=\left|\langle l|e^{i\frac{\phi_{c+}(x)}{2}}|l'\rangle\right|^2\, ,
\end{equation}
where $\phi_{c+}(x)$ is defined in Eq.~(\ref{eq:phicplus}), is a generalized FC factor, given by the overlapping of the wavefunctions of the molecular states $|l\rangle$ (with $4\kappa$ electrons )and $|l'\rangle$ (with $4\kappa+1$ electrons)~\cite{Franck,Condon}. Finally,
$f_{\eta}(E)=\left[1+e^{E/k_{B}T_{\eta}}\right]^{-1}$ is the Fermi
function for lead $\eta$ at temperature $T_{\eta}$ ($T_{S}=T+\delta T$, $T_{D}=T$),
\begin{equation}
\bar{\epsilon}=E_{c+}\left(4\kappa+\frac{1}{2}-N_{g}\right)+\frac{3}{8}\omega_{1}\, ,\label{eq:resen}
\end{equation}
and $\Omega_{ll'}=\Omega_{0}(l'-l)$. {Note that in order to derive Eqns.~(\ref{eq:R1},\ref{eq:R2}) we have assumed a symmetric voltage drop on both tunnel barriers. This choice is not restrictive for the present work as results in the linear transport regime do not depend on the particular choice adopted.}\\

\noindent The {\em position-dependent} FC factors can be explicitly evaluated as
\begin{equation}
F_{ll'}(x)=e^{-\lambda^{2}(x)}\left[\lambda(x)\right]^{2|l'-l|}\frac{l_{<}!}{l_{>}!}\left[L_{l_{<}}^{|l'-l|}\left(\lambda^{2}(x)\right)\right]^{2}\label{eq:FC}
\end{equation}
with
$l_{<}=\mathrm{min}\{l,l'\}$, $l_{>}=\mathrm{max}\{l,l'\}$,
$L_{p}^{q}(x)$ the Laguerre polynomials and
\begin{equation}
\lambda(x)=\sqrt{\frac{\omega_{1}}{\Omega_{0}}\left[\sum_{\nu\geq1}k_{\nu 0}\sin\left(\frac{\pi\nu
    x}{L}\right)\right]^{2}+\frac{\Omega_{0}}{\omega_{1}}\left[\sum_{\nu\geq1}\frac{k_{\nu 0}}{\nu}\cos\left(\frac{\pi\nu
    x}{L}\right)\right]^{2}}\label{eq:lambdaofx}
\end{equation}
a {\em position-dependent CNT-molecule coupling parameter}, whose explicit form depends on the diagonalization discussed in the previous section.\\
\noindent The coupling $\lambda(x)$ determines the behaviour of the
tunneling rates and allows the possibility to trigger transitions
between different molecular vibration state via the FC
factors $F_{ll'}(x)$. When $\lambda(x)\ll 1$, one has $F_{ll'}(x)\approx\delta_{l,l'}$ and tunneling events are not able to trigger transitions among different vibrational states. On the other hand, $F_{ll'}(x)\neq 0$ for several different values of $l$ and $l'$ when $\lambda(x)\neq 0$. The larger $\lambda(x)$, the wider is the jump $|l'-l|$ allowed by the FC factors, while transitions with $l\approx l'$ and/or small $l,l'$ are strongly suppressed. Where $\lambda(x)$ is sizeable, the transport properties of the CNT are then drastically influenced.\\
 
\noindent In the linear regime, the
strategy is to expand and solve Eq.~(\ref{eq:master}) to linear order
in $\delta V$ and $\delta T$ and then plug the solution into the
current expression Eq.~(\ref{eq:current}), again
retaining only the leading (linear) terms
\begin{equation}
	I\approx G_{V}\delta V+G_{T}\delta T\, .
\end{equation}
 The procedure is lengthy but straightforward~\cite{Furusaki,Koch} and shall be omitted here. We
just quote the final results
\begin{eqnarray}
G_{V}=G_{0}^{(V)}f(-\bar{\epsilon})\sum_{l,l'}e^{-\beta
  l\Omega_{0}}f\left[\bar{\epsilon}+\Omega_{ll'}-k_{B}T\log(4)\right]\chi_{l,l'}(x)\, ,\label{eq:condV}\\
G_{T}=G_{0}^{(T)}f(-\bar{\epsilon})\sum_{l,l'}e^{-\beta
  l\Omega_{0}}\left[\bar{\epsilon}+\Omega_{ll'}\right]f\left[\bar{\epsilon}+\Omega_{ll'}-k_{B}T\log(4)\right]\chi_{l,l'}(x)\, .\label{eq:condT}
\end{eqnarray}
Here, we have introduced {$G_{0}^{(V)}=\beta e^{2}\Gamma_{0}^{(D)}\left(1-e^{-\beta\Omega_{0}}\right)/2$,
$G_{0}^{(T)}=\beta^{2}e
k_{B}\Gamma_{0}^{(D)}\left(1-e^{-\beta\Omega_{0}}\right)/2$},
$\beta^{-1}=k_{B}T$, 
$f(E)=\left[1+e^{\beta E}\right]^{-1}$ and
\begin{equation}
\chi_{l,l'}(x)=\frac{4AF_{ll'}(x)F_{ll'}(x_{D})}{4F_{ll'}(x)+AF_{ll'}(x_{D})}\, ,
\end{equation}
with $A=\Gamma_{0}^{(D)}/\Gamma_{0}^{(S)}$ the barriers asymmetry
($A\gg 1$ for a typical STM setup). The logarithmic factors in Eq.~(\ref{eq:condV}) and Eq.~(\ref{eq:condT}) stem from the CNT spin and valley degeneracy and provide a shift of the peaks of $G_{V}$ and $G_{T}$ linear in temperature. We note here that the above results for $G_{V}$ and $G_{T}$ in the {\em linear} regime are identical to those which would have been obtained assuming the CNT and vibrational degrees of freedom fully relaxed~\cite{Koch}.
\section{Parameters and Results}
\label{sec:sec3}
Before discussing our results, let us review the parameters upon which
our model holds.\\
\noindent The CNT-molecule coupling parameter $\lambda(x)$ in Eq.~(\ref{eq:lambdaofx}) is influenced by the dimensionless parameter $\eta=\omega_{0}/\omega_{1}$ and to the
dimensionless coupling strength $\xi$ defined in Eq.~(\ref{eq:xidef}). For
librational modes one can estimate a typical frequency of the order of
$1\div 4$ Thz with $\omega_{0}\approx 0.6\div2.4\ m\mathrm{eV}$, while {\em for a weakly interacting CNT ($g\approx 1$)} $\omega_{1}\approx 2\ \mathrm{meV}$  assuming $L\approx 1\
\mu\mathrm{m}$, which yields $\eta\approx 0.2$. For shorter CNTs, and/or for lower values of $g$ (stronger interactions), $\eta$
decreases accordingly. {In general, the larger the mismatch between the molecule and plasmon frequency - and thus the smaller $\eta$ - the weaker are the effects of the CNT-molecule coupling~\cite{noi}.} The interaction strength $\xi$ in turn depends
on the electron-molecule interaction strength $V_{0}$, the CNT length and the molecule mass as well as on its vibrational frequency. Typical values for $V_{0}$ are in the range
$1\div40\ \mathrm{eV}\cdot\mathrm{m}$~\cite{couppap}. For the CNT, we assume here
an average length $L\approx200\ \mathrm{nm}$ and we consider fairly
light molecules (such as NH$_{3}$) with average molar mass of about
$20\ \mathrm{g/mol}$. This allows to estimate
$0.01\leq\xi\leq0.2$.\\
\noindent Estimates for the interaction range $w$ in Eq.~(\ref{eq:yukawa}) may be given on a
phenomenological basis only. We expect it to be
larger than the Luttinger cutoff length, thus satisfying
$\alpha<w<L$. In the following, we will consider values in the range $w\approx L/10$.\\
\noindent The above estimates satisfy the condition in Eq.~(\ref{eq:nowb}), ensuring the stability of the system and the validity of the proposed model.\\

\begin{figure}[htbp]
\begin{center}
\includegraphics[width=10cm,keepaspectratio]{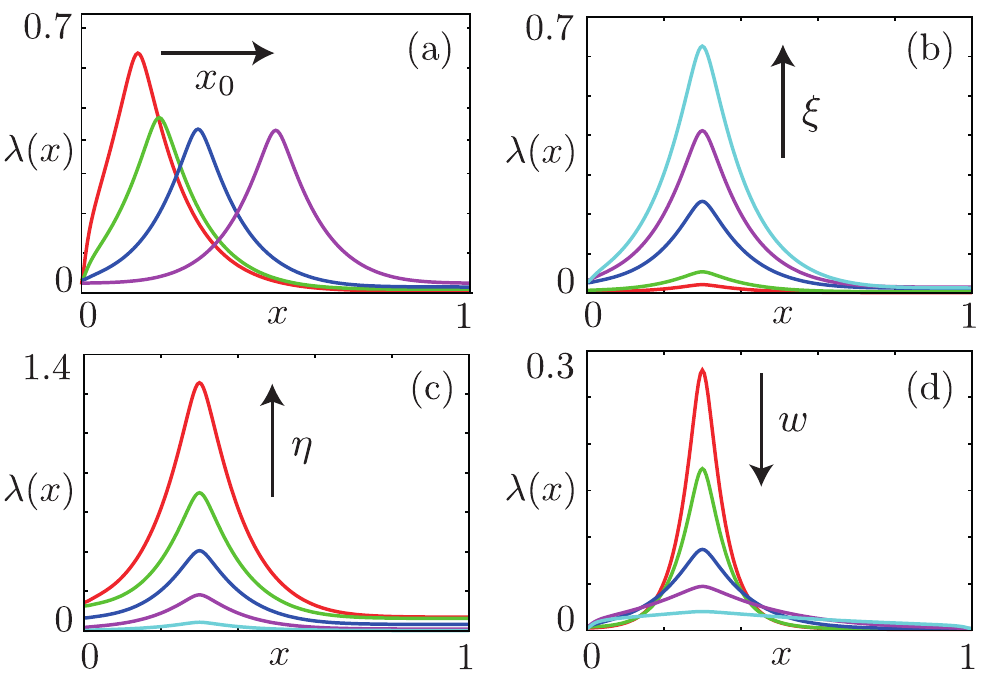}
\caption{Behaviour of $\lambda(x)$ as a function of the position $x$ (units $L$) of the STM tip along the CNT, for:\\
\noindent (a) different values of the molecule position $x_{0}/L$: 0.125
  (red), 0.2 (green), 0.3 (blue), 0.5 (purple). Here, $\xi=0.15$,
  $\eta=0.25$ and $w=L/10$;\\
\noindent (b) Different
  values of $\xi$: 0.01(red), 0.05 (green), 0.1 (blue), 0.15 (purple),
  0.2 (cyan). Here, $x_{0}/L=0.3$, $\eta=0.25$ and $w=L/10$;\\ 
\noindent (c) Different values of $\eta$: 0.05 (cyan), 0.2
  (purple), 0.4 (blue), 0.6 (green), 0.8 (red). Here, $x_{0}/L=0.3$,
  $\xi=0.1$ and $w=L/10$;\\
\noindent (d) Different
  values of $w/L$: 0.1 (red), 0.2 (green), 0.3 (blue), 0.4
  (purple), 0.5 (cyan). Here, $x_{0}/L=0.3$, $\xi=0.1$, and
  $\eta=0.1$.\\
\noindent In all panels, $\alpha=L/30$ {and $g=1$}. The arrows denote the direction of increasing parameters.}
\label{fig:fig4}
\end{center}
\end{figure}

\noindent Figure~\ref{fig:fig4}(a) shows the effective coupling
parameters $\lambda(x)$ as a function of $x$, for different positions
of the molecule attached to the CNT. As a general feature, we observe
that $\lambda(x_{D})\approx 0$ unless $x_{0}\approx x_{D}$, i.e. when the molecule is very close to the drain contact (a rather
peculiar situation). This means that in general tunneling events at
the lateral contact will be rather unefficient in triggering molecular
vibrations. On the other hand, the STM tip - being able to scan the
CNT length - can excite oscillations in the molecule when it is
located in proximity of the latter. Indeed, the curves in
Fig.~\ref{fig:fig4}(a) always display a maximum of $\lambda(x)$
located at $x=x_{0}$. This already suggests that the STM tip can be a
valuable tool to investigate the properties of the molecule. This will
be confirmed by
the results on the thermopower shown later in this section. The value of $\lambda(x_{0})$ exhibits a slight increase as $x_{0}\to 0$ (or $x_{0}\to L$). This is due to the spatial behaviour of the long-wave electron density: indeed one finds that in the parameters regime considered in this work, when $x_{0}$ deviates from the CNT center, $\lambda(x)$ is dominated by terms with $\nu\approx 1$ in Eq.~(\ref{eq:lambdaofx}). Inspecting Eq.~(\ref{eq:gamma3}) one observes that these terms are all large and in phase when $x_{0}$ is near the CNT borders.\\

\noindent Given the relatively large range of variation of the
different system parameters, it is worthwile to investigate their
effects on $\lambda(x)$. Figure~\ref{fig:fig4}(b) shows $\lambda(x)$
for different value of $\xi$. As it may be expected, since $\xi$
directly parameterizes the strength of the molecule-CNT coupling,
$\lambda(x)$ increases (decreases) when $\xi$ increases (decreases). A
similar behaviour is observed when $\eta$ is varied - as shown in
Fig.~\ref{fig:fig4}(c). The coupling between the molecule and the CNT
is more efficient if $\eta$ increases, i.e. when the vibrational frequency and the plasmon frequency are
closer and thus more resonant. Finally, Fig.~\ref{fig:fig4}(d) shows the dependence of
$\lambda(x)$ on the width of the molecule-CNT interaction potential
$w$: the width of the effective coupling closely follows $w$ with its maximum
scaling as $w^{-1}$, in agreement with the behaviour of $V(x)$ in Eq.~(\ref{eq:yukawa}).\\

\begin{figure}[htbp]
\begin{center}
\includegraphics[width=10cm,keepaspectratio]{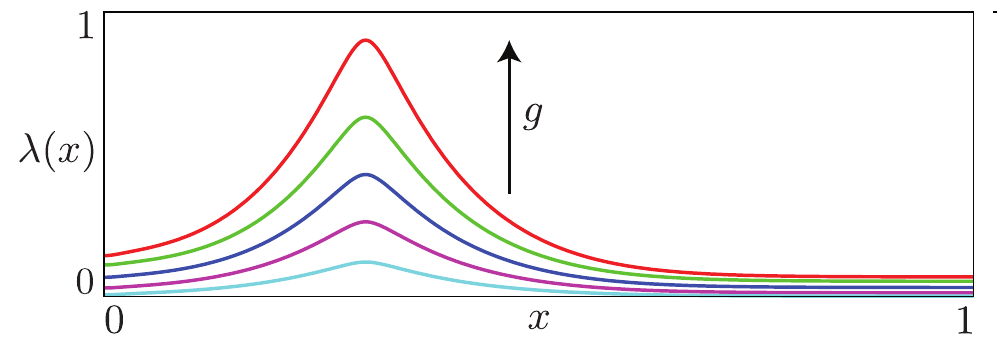}
\caption{{Behaviour of $\lambda(x)$ as a function of the position $x$ (units $L$) of the STM tip along the CNT, for different values of the Luttinger interaction parameter $g=$: 1
  (red, corresponding to $\eta=0.7$), 0.8 (green, $\eta=0.56$), 0.6 (blue, $\eta=0.42$), 0.4 (purple, $\eta=0.28$), 0.2 (cyan, $\eta=0.14$). Here, $L=875\ \mathrm{nm}$, $\omega_{0}=2\ \mathrm{Thz}$, $x_{0}=0.3\ L$, $\xi=0.15$, and $w=L/10$, and $\alpha=L/30$. The arrows denote the direction of increasing $g$ and thus of {\em decreasing} interaction strength.}}
\label{fig:fig5}
\end{center}
\end{figure}

\noindent{
\noindent Figure~\ref{fig:fig5} shows the effects of Coulomb interactions on the coupling $\lambda(x)$. Clearly, as $g\to 0$ (from red to cyan curve) the coupling gets suppressed. This fact is due to the decrease of the parameter $\eta\propto g$. We want to stress here that additional effects induced by the presence of strong Coulomb interactions, such as the formation of a Wigner molecule in the CNT, are ignored since we assume the CNT to have a large number of particles $N_0$ with a corresponding Wigner wavelength shorter than the width $w$ of the molecule interaction potential, see Sec. 2. The fact that the CNT-molecule coupling is larger for weakly interacting systems constitutes an advantage from the experimental point of view, due to almost unavoidable presence of surrounding screening metalizations.}\\

\begin{figure}[htbp]
\begin{center}
\includegraphics[width=10cm,keepaspectratio]{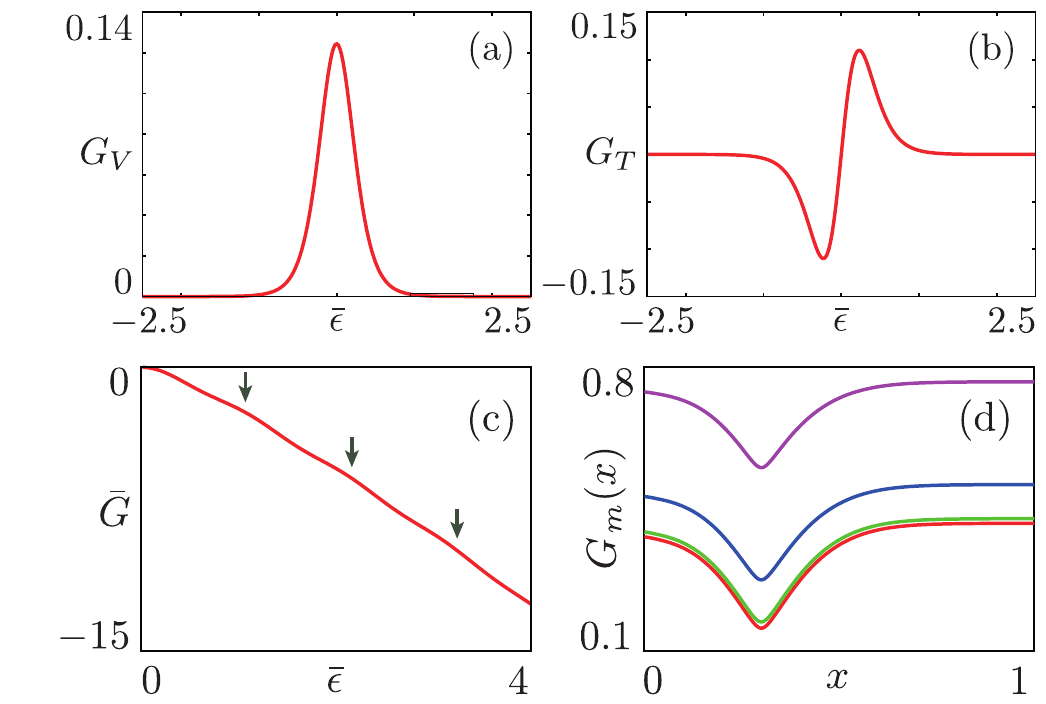}
\caption{{(a) Plot of $G_{V}$ (units $G_{0}^{V}$) as a function of
  $\bar{\epsilon}$ (units $\Omega_{0}$); (b) Plot of $G_{T}$ (units
  $G_{0}^{(T)}$) as a function of $\bar{\epsilon}$ (units
  $\Omega_{0}$). (c) Plot of $\bar{G}=\log\left[G_{V}/G_{V}(\bar{\epsilon}=0)\right]$ as a function of $\bar{\epsilon}$ (units
  $\Omega_{0}$). Arrows denote the position of the exponentially suppressed vibrational sidebands induced by the coupling between the CNT and the molecule. (d) Plot of the conductance maximum $G_{m}(x)$
  normalized to $G_{0}^{(V)}$ as a function of $x$ (units $L$) for
  different temperatures $T$ (units $\Omega_{0}/k_{B}$): 0.05 (red),
  0.1 (green), 0.25 (blue), 0.5 (purple). Parameters for all panels:
  $\xi=0.1$, $\eta=0.6$, $w=L/10$, $A=100$ and $\alpha=L/30$. In
  panels (a,b), $k_{B}T=0.1\Omega_{0}$ and $x=x_{0}=0.3L$.}}
\label{fig:fig6}
\end{center}
\end{figure}

\noindent Let us now turn to the linear transport properties. Before
investigating the thermopower $S$, it is useful to study the charge
and thermal conductances. Figure~\ref{fig:fig6}(a,b) shows the typical
behaviour of $G_{V}$ and $G_{T}$ as a function of the resonant energy $\bar{\epsilon}$, see Eq.~(\ref{eq:resen}). The linear charge conductance exhibits a peak
centered around the main resonance value $\bar{\epsilon}=0$, while the
thermal conductance shows a kink centered in the same position. These
features are due to the contributions with $l=l'\approx 0$ in
Eq.~(\ref{eq:condV}) and Eq.~(\ref{eq:condT}). Also the terms with $l\neq l'$ and possibly $l,l'\neq 0$,
due to the excitation of the molecule vibrational modes, contribute to
the above quantities and {lead in principle to satellite peaks red- and
blue-shifted with respect to the main resonance, located around
$\bar{\epsilon}\approx n\Omega_{0}$ ($n$ an integer). This would in principle allow to directly extract the vibrational frequency of the molecule from a measurement of $G_{V}$. These peaks, however, are not visible: in the regime $k_{B}T\ll\Omega_{0}$
they are exponentially suppressed, while at higher
temperature they are subdued by the tails of the
thermally-broadened central peak. Figures~\ref{fig:fig6}(a,b) indeed show no detectable trace of such sidebands. Figure~\ref{fig:fig6}(c) shows a plot of
\begin{equation}
\bar{G}=\log{\left[\frac{G_{V}}{G_{V}(\bar{\epsilon}=0)}\right]}\, .
\end{equation}
For a single conductance peak, one would expect a featureless exponential decay of $G_{V}$. However, the plot shows faint oscillations at the vibron resonance positions $\bar{\epsilon}=n\Omega_{0}$. Such oscillations are virtually undetectable in the linear plot of the conductance and extremely unlikely to be observed in experiment. Also $G_{T}$ displays analogous features (not shown here).}\\
\noindent Let us now analyze in more details the main peak of $G_{V}$, concentrating on the low
temperature regime $k_{B}T<\Omega_{0}$. Concentrating around $\bar{\epsilon}\approx 0$ and retaining only the terms $l=l'$ in Eq.~(\ref{eq:condV}) one obtains
{
\begin{equation}
  G_{V}\approx \frac{G_{0}^{(V)}}{\cosh{\left(\frac{\beta\bar{\epsilon}}{2}\right)}\cosh{\left[\frac{\beta\bar{\epsilon}-log(4)}{2}\right]}}\sum_{l}e^{-l\beta\Omega_{0}}\frac{AF_{ll}(x)}{A+4F_{ll}(x)}\, ,\label{eq:approxGv}
\end{equation}}
where we have exploited the fact that $\lambda(x_{D})\approx 0$. The conductance exhibits a maximum for $\bar{\epsilon}=0$ with the well-known $T^{-1}$ power-law scaling through the factor
$G_{0}^{(V)}$. When $A\ll 1$ the conductance is essentially independent of the STM tip position, since conductance is dominated by the lateral conductance only. This case however is not very interesting from the point of view of a realistic STM setup, where $A\gg 1$. In this case, one can perform the summation in Eq.~(\ref{eq:approxGv}) and obtain an analytic expression for the {\em space-dependent} conductance maximum 
{
\begin{equation}
G_{m}(x)=\frac{8}{9}\frac{G_{0}^{(V)}}{1-e^{-\beta\Omega_{0}}}\exp\left[-\frac{\lambda^2(x)}{1-e^{-\beta\Omega_{0}}}\right]\, .\label{eq:maxshape}
\end{equation}}
The above expression is especially simple for $k_{B}T\ll\Omega_{0}$ when one finds {$G_{m}(x)\approx (8/9)G_{0}^{(V)}e^{-\lambda^2(x)}$}.
A plot of $G_{m}(x)$ is shown in Fig.~\ref{fig:fig6}(c) as a
function of $x$ for different temperatures. As expected by Eq.~(\ref{eq:maxshape}), it exhibits a dip in the location where the molecule sits. Therefore, measuring the
maximum of $G_{V}$ as a function of the STM
position allows to extract $\lambda(x)$ and ultimately reveals the
position of the molecule along the CNT. 

\begin{figure}[htbp]
\begin{center}
\includegraphics[width=10cm,keepaspectratio]{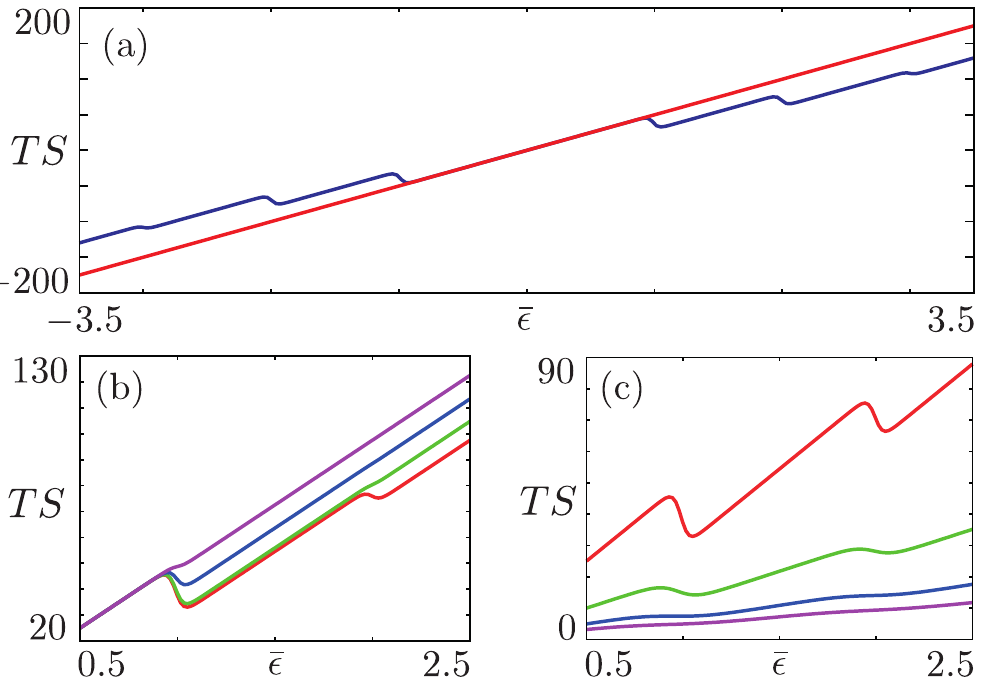}
\caption{Plot of $TS$ (units $\Omega_{0}/e$) as a function of
  $\bar{\epsilon}$ (units $\Omega_{0}$) for: (a) the STM tip at
  $x=0.8\ L$ (away from the molecule, red) and at $x=x_{0}$ (ontop of the
  molecule, blue) with $k_{B}T=0.02\Omega_{0}$ and $A=100$; (b) for the tip at $x=x_{0}$ and different barrier
  asymmetries: $A=100$ (red), $A=50$ (green), $A=10$ (blue), $A=2$
  (purple) with $k_{B}T=0.02\Omega_{0}$; (c) for the tip at $x=x_{0}$ and different
  temperatures $T$ (units $\Omega_{0}/k_{B}$): 0.05 (red), 0.1
  (green), 0.15 (blue), 0.2 (purple) with $A=100$. Other parameters for all panels:
  $\xi=0.1$, $\eta=0.6$, $w=L/10$, $\alpha=L/30$ and $x_{0}=0.3L$.}
\label{fig:fig7}
\end{center}
\end{figure}

\noindent Finally, we turn to the thermopower $S$, shown as a function
of $\bar{\epsilon}$ in Fig.~\ref{fig:fig7}(a). When the tip is away of
the molecule, red line, it shows a featureless linear trend as a
function of $\bar{\epsilon}$, typical of its behaviour around a
Coulomb blockade oscillation~\cite{BeenakkerThermo}. We stress here that we are concentrating in a {\em low-energy} regime in which the electronic collective excitations of the CNT are effectively frozen out, leaving no signature on the thermopower $S$. On the other
hand, when the tip is located at $x=x_{0}$, blue curve, a
non-monotonic behaviour as a function of $\bar{\epsilon}$ is
shown. Each sawtooth-like dip is located at
$\bar{\epsilon}=n\Omega_{0}$ ($n$ an integer) and signals the
triggering of vibrational excitations in the molecule, in analogy to
the behaviour observed in nano-electromechanical
systems~\cite{Koch}. The thermopower $S$ is therefore able to
detect the presence of a molecule attached to the CNT, by displaying
vibrational sidebands at $\bar{\epsilon}=n\Omega_{0}$. We remind here that usually, such vibrational sidebands are detected in the {\em non-linear} transport regime. {Since both $G_{T}$ and $G_{V}$ display the same exponential suppression for the vibrational sidebands, their ratio becomes insensitive to it. As a result, $S$ displays clear signatures of quantities not directly accessible in the direct measurements of the conductance~\cite{BeenakkerThermo}.} In analogy with $G_{V}$, the visiblity of this effect is strongly
influenced by the asymmetry of tunnel barriers $A$, see
Fig.~\ref{fig:fig7}(b). As $A$ decreases, the sidebands get weaker and
eventually vanish altogether when $A\approx 1$. The STM setup,
therefore, operates in the correct regime to exploit the information
contained into $S$. Also thermal effects tend to smear out the above
features, as shown in Fig.~\ref{fig:fig7}(c): in
order to obtain the maximum contrast one needs $k_{B}T\ll\Omega_{0}$.\\

\begin{figure}[htbp]
\begin{center}
\includegraphics[width=10cm,keepaspectratio]{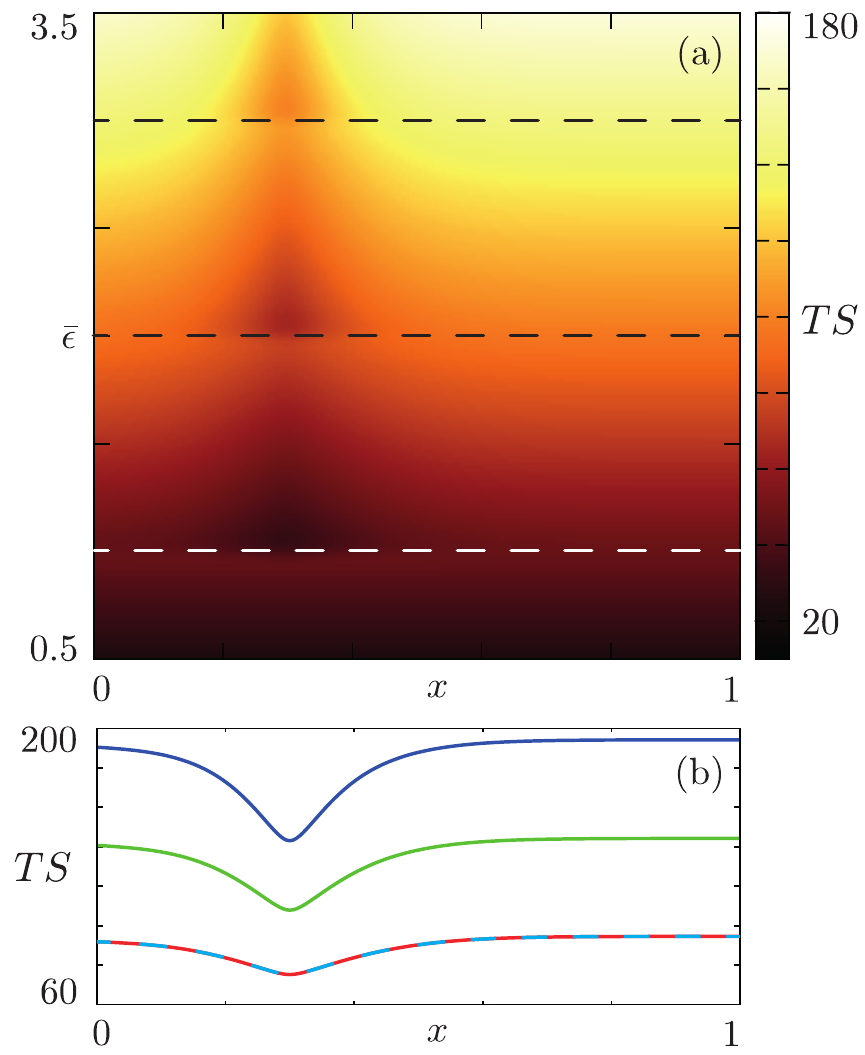}
\caption{(a) Colormap plot of $TS$ (units $\Omega_{0}/e$) as a
  function of $x$ (units $L$) and $\bar{\epsilon}$ (units
  $\Omega_{0}$). The dashed lines correspond to the values of
  $\bar{\epsilon}$ chosen for the next panel. (b) Plot of $TS$ (units
  $\Omega_{0}/e$) as a function of $x$ (units $L$) for different
  values of $\bar{\epsilon}$: $\bar{\epsilon}=\Omega_{0}$ (red),
  $\bar{\epsilon}=2\Omega_{0}$ (green), $\bar{\epsilon}=3\Omega_{0}$
  (blue). The dashed curve represents the analytic approximation in Eq.~(\ref{eq:anapS}). In all panels: $\xi=0.1$,
  $\eta=0.6$, $w=L/10$, $A=100$, $\alpha=L/30$, $k_{B}T=0.1\Omega_{0}$, and $x_{0}=0.3L$.}
\label{fig:fig8}
\end{center}
\end{figure}

\noindent The thermopower, however, encodes information about the {\em
  position} of the molecule as well. Figure~\ref{fig:fig8}(a) shows a
colorscale map of $S$ as a function of the STM tip position $x$ and
$\bar{\epsilon}$. Following the graph along vertical lines, one
obtains the $S$ vs. $\bar{\epsilon}$ plots discussed above. For $x$ away from $x_{0}$ an increase of $S$ as a function of
$\bar{\epsilon}$ is observed, while features develop for $x\approx
x_{0}$. Figure~\ref{fig:fig6}(b) shows curves of $S$ as a function of
the tip position $x$ for the resonance conditions
$\bar{\epsilon}=n\Omega_{0}$ ($n=1,2,3$). Each of these curves
exhibits a dip located at $x_{0}$, thus confirming that the thermopower is a tool able to detect the location of the attached molecule.\\
\noindent In order to be more quantitative, we study the dip of $S$ around
$\bar{\epsilon}\approx\Omega_{0}$, considering for definiteness the
low-temperature regime $k_{B}T\ll\Omega_{0}$. Only terms with $l'=0$ and $l=0,1$ must then be retained in Eq.~(\ref{eq:condV}) and Eq.~(\ref{eq:condT}). Neglecting the logarithmic factors, irrelevant at small temperatures, one gets
\begin{equation}
S\approx\frac{\Omega_{0}}{e T}\frac{2\chi_{0,0}(x)}{2\chi_{0,0}(x)+\chi_{1,0}(x)}\approx\frac{\Omega_{0}}{e T}\frac{2}{2+\lambda^{2}(x)}\, ,\label{eq:anapS}
\end{equation}
for $A\gg 1$. The plot of this expression is shown in
Fig.~\ref{fig:fig8}(b) as a dashed dot, in agreement with full numerics. Similar approximations can be obtained for the
resonances at higher energy. Their expressions hovever are too
involved to be reported here.\\
Thus, a measurement of $S$ as the tip scans the CNT allows to extract $\lambda(x)$ and
confirms that thermopower is a powerful tool to extract information about the molecule and on its location.
\section{Conclusions}
\label{sec:sec4}
In this paper we have considered the transport properties of a CNT
capacitively coupled to a molecule vibrating along its librational
modes. The CNT-molecule coupling has been diagonalized exactly extracting the new eigenmodes of the system. Transport
in the presence of an STM tip scanning the CNT length has been
considered. Employing a master equation approach in the sequential
tunneling regime, analytic expressions for the linear charge and
thermal conductances have been obtained, allowing to study the behaviour of the thermopower. The transport properties depend on the tip
position and are characterized by the appearance of a
position-dependent coupling to the molecule, with a maximum peaked at
its position.\\

\noindent The linear charge conductance of the system allows to
extract information about the above position-dependent
coupling. However, it is not a sensitive probe for the vibrational
spectrum of the molecule. On the other hand, the thermopower shows
features corresponding to resonances with the molecular vibrations,
allowing thus to extract the molecular vibration frequency. These
features also strongly depend on the position and are shown to be
directly related to the effective CNT-molecule coupling thus allowing
to locate the molecule along the CNT.\\

\noindent\textit{Acknowledgments.} Financial support by the EU- FP7
via ITN-2008-234970 NANOCTM is gratefully acknowledged.

\end{document}